\documentclass[aps,prl,a4paper,reprint,amsmath,amssymb,floatfix,showpacs]{revtex4-1}
 \usepackage[T1]{fontenc}
 \usepackage{amsmath}
 \usepackage{graphicx}
 \usepackage{SIunits}
 \usepackage{ulem}
 \usepackage{braket}
 \usepackage{color}
 \usepackage{natbib}

\renewcommand{\vec}[1]{\ensuremath{\boldsymbol{#1}}}
\newcommand{\abs}[1]{\ensuremath{|#1|}}
\newcommand{\Q}{\ensuremath{Q}}

\begin{document}

 \title{Statistical theory of a quantum emitter strongly coupled to Anderson-localized modes}

 \date{\today}
 \author{Henri Thyrrestrup}
 \author{Stephan Smolka} \thanks{Present address: Institute of Quantum Electronics, ETH Z\"urich, 8093 Z\"urich, Switzerland}
  \affiliation{Niels Bohr Institute, University of Copenhagen, Blegdamsvej 17, Dk-2100 Copenhagen, Denmark}
    \author{Luca Sapienza} \thanks{Present address: School of Engineering and Physical Sciences, Heriot-Watt University, Edinburgh EH14 4AS, UK}
 \author{Peter Lodahl}
 \email{lodahl@nbi.ku.dk}
 \affiliation{Niels Bohr Institute, University of Copenhagen, Blegdamsvej 17, Dk-2100 Copenhagen, Denmark}

 \begin{abstract}
A statistical theory of the coupling between a quantum emitter and Anderson-localized cavity modes is presented based on a dyadic Green's function formalism. The probability of achieving the strong light-matter coupling regime is extracted for an experimentally realistic system composed of InAs quantum dots embedded in a disordered photonic crystal waveguide. We demonstrate that by engineering the relevant parameters that define the quality of light confinement, i.e. the light localization length and the loss length, strong coupling between a single quantum dot and an Anderson-localized cavity is within experimental reach. As a consequence of disorder-induced light confinement provides a novel platform for quantum electrodynamics experiments.
 \end{abstract}


 \maketitle

The ability of controlling the coupling between a single emitter and a photon grants access to a rich set of quantum optical phenomena \cite{Microcavities} and provides a route towards quantum networks for quantum information technology \cite{Kimble}. The coupling strength can be enhanced by placing the emitter in a small optical cavity that traps the emitted photon \cite{Vahala}. As a result, the spontaneous emission rate is enhanced through the Purcell effect and if the coupling strength exceeds the cavity loss rate, the strong coupling regime is reached where coherent oscillations between photon and emitter induce quantum entanglement \cite{JMG}. In solid state systems, strong coupling has been achieved, e.g., by embedding semiconductor quantum dots in highly engineered photonic crystal cavities with small mode volumes and high \Q-factors \cite{Yoshie}. However, the fabrication of such cavities requires nanoscale accuracy and is therefore inherently sensitive to unavoidable disorder introduced in the fabrication process.

An alternative route to efficient light confinement based on disordered photonic nanostructures has recently proven to be very promising. In such systems the light propagation is strongly modified by multiple scattering, which can lead to the spontaneous formation of Anderson-localized modes. These modes are robust against fabrication imperfections and prevail even after averaging over all configurations of disorder \cite{Pin Sheng}.Disordered photonic crystal waveguides are well suited for investigating Anderson localization since the light propagation is primarily one-dimensional (1D) and localized modes are formed when the sample length is sufficiently long. Impressive \Q-factors as high as $600.000$ have been reported in such disorder-induced modes \cite{VollmerPRL,Vollmer} that subsequently were proven to be due to Anderson localization by carrying out the proper configuration average \cite{David,Science paper}.

Recently, Anderson-localized modes were shown to enable very pronounced cavity quantum electrodynamics effects by probing the dynamics of single quantum dots embedded in disordered photonic crystal waveguides \cite{Science paper}. Here we present a theoretical model for photon-matter interaction of a single quantum emitter in a 1D disordered medium in the Anderson-localized regime. In a multiple scattering medium, light propagation and emission is determined by a statistical process giving rise to a distribution of photon-matter coupling rates. Two universal parameters fully characterize the propagation of light in a 1D random medium: the localization length scaled to the sample length $\xi/L$ and the loss length scaled to localization length $l/\xi$  \cite{Pin Sheng,Deych2001}, while the emission is non-universal in the sense that it also depends on the local correlation length of the medium in the presence of the emitter \cite{Skipetrov,Tiggelen}. We derive a distribution of optical modes in the Anderson-localized regime of a 1D disordered medium and based on a dyadic Green's function formalism extract the associated distribution of photon-matter coupling coefficients. From this model we calculate the probability of strong coupling of a single quantum dot to an Anderson-localized cavity mode and the dependence on localization length and loss length. Considering realistic parameters obtained in disordered photonic crystal waveguides \cite{Stephan} we conclude that the experimental observation of strong coupling of a quantum dot to an Anderson-localized mode is within experimental reach.

\begin{figure}[tb]
    \centering
    \includegraphics[width=0.9\linewidth]{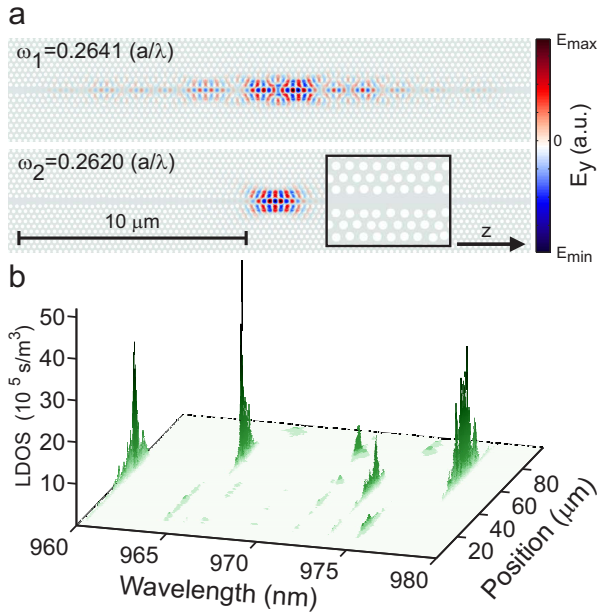}
    \caption{(color online). (a) 2D finite-difference time-domain calculations of the $y$-component of the electromagnetic field (color scale) in a disordered photonic crystal waveguide. We have used realistic parameters for a GaAs photonic crystal waveguide with lattice parameter $a=\unit{260}{\nano\meter}$, air-hole radius $r=0.29a$, and a refractive index of $n=2.76$, the latter being the effective refractive index of a \unit{150}{\nano\meter} thick GaAs membrane. Disorder is introduced by displacing the position of the holes in the three rows on each side of the waveguide with a Gaussian random distribution with standard deviation of $\sigma = 0.03a$. Inset: Closeup view of the geometry of the disordered waveguide (b) Local density of state (LDOS) for a single realization of disorder using a stack of layers with refractive indices homogenously distributed in the interval $3.45 \pm 0.70$ leading to $\xi = \unit{15}{\micro\meter}$ for $L=\unit{100}{\micro\meter}$, corresponding to realistic values from experiments on PhC waveguides \cite{Stephan}}.
    \label{fig:LDOS_struct}
\end{figure}

In a disordered W1 photonic crystal (PhC) waveguide membrane light propagation is effectively 1D since it is confined to the membrane by total internal reflection and to the waveguide by the 2D PhC band gap. Disorder introduces multiple random scattering of the Bloch waves and the localization length, $\xi$, denotes the average distance between scattering events. The criterion for Anderson localization is $\xi < L$ leading to an exponential decay of the light intensity emitted by a source after ensemble averaging over all realizations of disorder \cite{Baron}. Examples of calculated electric field profiles of Anderson-localized modes in a PhC waveguide are shown in Fig.~\ref{fig:LDOS_struct}(a) where disorder is introduced in the waveguide by random displacements of the hole positions in the three rows on each side of the waveguide. Anderson-localized modes are found to appear only in a narrow frequency interval near the cutoff of the waveguide mode where the density of states is high \cite{David,Toke-PRL,Thyrrestrup}. The localization length, loss length, and sample length are three essential parameters of a 1D multiple scattering system that enter into the model for light-matter coupling. In the following we will use physically realistic parameters and estimate the coupling of single quantum dot emitters to disordered PhC waveguides.

 The local light-matter coupling strength is determined by the local density of optical states (LDOS) that is calculated from the imaginary part of the frequency dependent dyadic Green's function \cite{NanoOptics}. The disordered waveguide consists of an $L=\unit{100}{\micro\meter}$ long stack of layers with randomly varying refractive indices along the $z$-direction with an average refractive index of $\bar n = 3.45$ and a layer thickness of \unit{10}{\nano\meter}. We consider the situation relevant for cavity QED experiments where the LDOS is dominated by localized quasi-modes and therefore can be expressed as $\rho(\vec r,\omega)=\abs{\vec{\tilde f}(x,y,\omega)}^2\tilde{\rho}(z,\omega)/A_{\text{eff}}(\omega)$, where $\tilde{\rho}(z,\omega)$ accounts for the spatial variation of the LDOS along the $z$ direction that is obtained from a self-consistent transfer-matrix method \cite{selfconsistent}, and $\omega$ is the optical frequency. The transverse effective area $A_{\text{eff}}(\omega)=\int \mathrm{d}x\mathrm{d}y\,\tilde n^2(x,y)\abs{\vec{\tilde f}(x,y,\omega)}^2\,$ is calculated from an ordered photonic crystal waveguide, where $\tilde n(x,y)$ is the refractive index and $\vec{\tilde f}(x,y,\omega)$ is the simulated transverse electric field profile \cite{MPB}, and both have been scaled to unity. The electric field profile is obtained from 3D band structure calculations and the weak dependence on $z$ is eliminated by averaging over one unit cell. All parameters used in the calculation of the PhC are listed in Fig.~\ref{fig:LDOS_struct}, and only the size of the refractive index fluctuations in the 1D multiple scattering model is varied giving rise to different localization lengths $\xi = 7.40 /(\Delta n)^2 \micro\meter$ for the $L=\unit{100}{\micro\meter}$ sample length and the particular form of disorder introduced in the present analysis.

\begin{figure}[tb]
    \centering
    \includegraphics[width=0.9\linewidth]{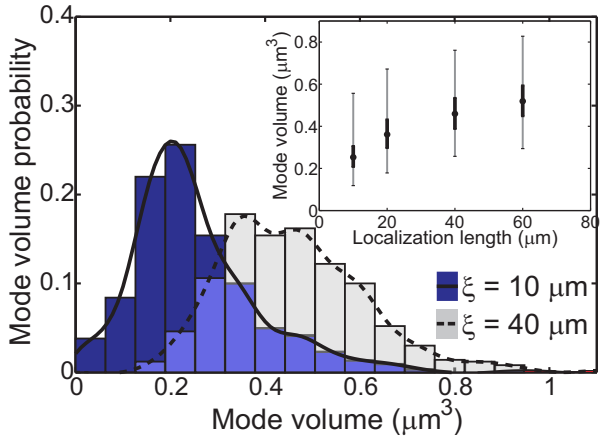}
    \caption{(color online). Histograms of the mode volume probability distribution calculated for 500 realizations of disorder and for two values of the localization length $\xi$. The transverse extension of the modes is obtained from the simulation of the PhC waveguide shown in Fig.~\ref{fig:LDOS_struct}(a). The solid and dashed lines are guides to the eye. Inset: Mean values and one and two standard deviations of the calculated mode volume distributions plotted as a function of localization length.}
    \label{fig:Vm_dist}
\end{figure}

An example of the spatial and spectral variations of the LDOS for a single configuration of disorder is displayed in Fig.~\ref{fig:LDOS_struct}(b), showing well-separated quasi-modes with large fluctuations in $\rho(\vec r)$ along the mode. The LDOS is sufficient to calculate the light-matter coupling, but a quasi-mode description is instructive for physical insight. It is obtained by fitting the LDOS by a sum of Lorentzians:
\begin{equation}
  \rho(\vec r,\omega) = \sum_i \rho_{0,i}(\vec r) \frac 1\pi \frac{\kappa_i/2}{(\omega-\omega_{i})^2+(\kappa_i/2)^2},
  \label{eq:LDOS}
\end{equation}
where each term of the sum describes a single quasi mode with resonance frequency $\omega_{i}$, photon decay rate $\kappa_i=\omega_{i}/Q_i$, quality factor $Q_i$, and amplitude $\rho_{0,i}(\vec r)$.
It is instructive to express the coupling strength in terms of the mode volume $V_i$ that enters in the Jaynes--Cummings model \cite{JMG}. For an emitter on resonance with an Anderson-localized mode $(\omega = \omega_i)$ and with a transition dipole moment oriented along the local electric field, the Purcell factor is $F_i(\vec r) = 6\pi c^3Q\rho_{0,i}(\vec r)/\omega_c^3 n(\vec r)$, and from that the effective mode volume $V_i = 1/\max_{\vec r}\{n^2(\vec r)\rho_{0,i}(\vec r)\}$ is defined. The distribution of mode volumes is subsequently calculated from an ensemble of 500 different disorder configurations for each value of the localization length and plotted in Fig.~\ref{fig:Vm_dist}. The distribution is found to shift to smaller mode volumes with decreasing localization length in agreement with the intuitive expectation that a short localization length leads to small mode volume and hence efficient cavity QED. For a localization length of $\xi=\unit{10}{\micro\meter}$ we extract mode volumes as small as $V_i \simeq 2.5 (\lambda/\bar n)^3,$ which is competitive with engineered PhC nano-cavities \cite{Noda}. In recent experiments using quantum dots to probe Anderson-localized modes a localization length below $\xi=\unit{10}{\micro\meter}$ has been extracted \cite{Stephan} and the ultimate lower bound on the localization length has not yet been established \cite{Lagendijk50Years}. This illustrates the very promising potential of efficient confinement of light with Anderson-localized modes.

\begin{figure}[tb]
    \centering
    \includegraphics[width=0.9\linewidth]{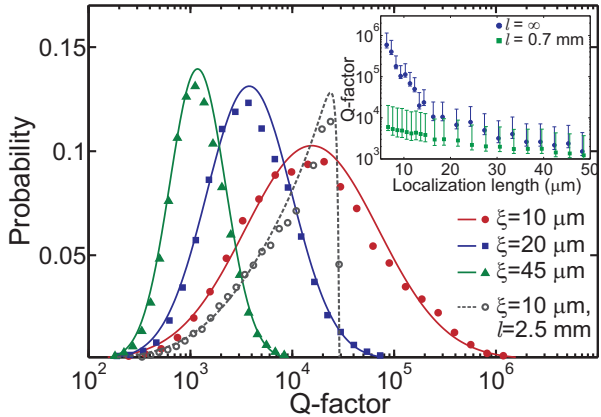}
    \caption{(color online). Normalized \Q-factor probability distributions for different localization lengths between $\xi=\unit{10-45}{\micro\meter}$ and without loss (filled symbols) with fits to log-normal distributions (solid lines), and for a loss length $l=\unit{2.5}{\milli\meter}$ and $\xi = \unit{10}{\micro\meter}$ (open circles) with a fit to a log-normal distribution modified to include loss \cite{Stephan}(dashed line). The sample length is $L=\unit{100}{\micro\meter}$. Inset: Mean value of the \Q-factor distributions (symbols) and the central 68\% of the \Q-factor distribution (bars), with and without loss (blue dots and green squares, respectively). }
    \label{fig:Q_dist}
\end{figure}

The $Q$-factor is the additional figure-of-merit determining the light-matter coupling strength. The statistics of the $Q$-factors is obtained from an ensemble of 8000 disorder configurations  for each localization length leading to the identification of approximately 60.000 Anderson-localized modes near the center of the waveguide and within the spectral range of $\lambda=\unit{970-980}{\nano\meter}.$
Figure~\ref{fig:Q_dist} plots the extracted $Q$-factor distributions for different localization lengths verifying that the $Q$-factors of Anderson-localized modes in 1D are described by log-normal distributions \cite{Pinheiro2008,Patra2003} (solid lines in Fig.~\ref{fig:Q_dist}). Due to the long tails of the log-normal distributions, a finite probability exists of having very large $Q$-factors that are much larger than the most probable value. The inset of Fig.~\ref{fig:Q_dist} shows that the average $Q$-factor increases super-exponentially with an increasing localization length, which illustrates the very sensitive dependence of the light-matter interaction on the localization length that is responsible for the pronounced cavity QED effects observed in experiments \cite{Science paper}.

In real PhC waveguides, loss primarily due to scattering of light out of the membrane structure limits the quality of the light confinement. Such a loss is included as a finite value of the imaginary part $n''$ of the refractive index $n'+i n''$ leading to the loss length $l=\lambda/2\pi n''$ and an associated loss \Q-factor $Q_{\text{loss}}'\pi/\lambda l$, which is valid in the limit where  loss can be treated perturbatively. This results in an effective $\Q$-factor $Q_{\text{eff}}^{-1} = Q^{-1} + Q^{-1}_{\text{loss}}$, where $Q_{\text{loss}}$ sets the limit of the highest \Q-factor that can be reached. An example is shown for $\xi=10\,\mu$m in Fig.~\ref{fig:Q_dist} (open circles) leading to a cut-off at $Q_{\text{loss}}=28000$ for the loss length of $l=\unit{2.5}{\milli\meter}$. The cut-off results in an interesting reshaping of the \Q-factor distribution since the highest \Q-factors from the loss-less case all accumulate in the part of the distribution with highest $Q$. The overall reduction in the \Q-factors in the presence of loss is reflected in the average \Q-factor that is displayed in the inset of Fig.~\ref{fig:Q_dist}.

We now turn to the statistic of the light matter coupling strength and derive the probability of obtaining strong coupling in disordered waveguides. By applying the decomposition into quasi-modes of Eq.~\eqref{eq:LDOS}, the threshold for reaching strong coupling for a single emitter that is resonant with a single Anderson-localized mode $i$, is given as
\begin{equation}
  \frac{Q_i^2 \rho_{0,i}(\vec r) }{\omega_c} > \frac{\varepsilon_0\hbar}{8 d^2}.
  \label{eq:scrongcoupling}
\end{equation}
The right hand side of Eq.~\eqref{eq:scrongcoupling} only contains physical constants and the dipole moment $d$ of the emitter while the left hand side contains all the optical properties of the localized modes. For identical emitters the probability for reaching the strong-coupling regime is therefore solely determined by the distributions of $Q_i$ and $\rho_{0,i}$ that are both linked to the localization length, as explored in the data of Fig.~\ref{fig:Vm_dist} and Fig.~\ref{fig:Q_dist}. We stress that a shortening of the localization length is beneficial in two ways since it simultaneously increases the $Q$-factor and decreases the mode volume, which is the underlying reason for the success of Anderson-localized cavities in cavity QED experiments.

\begin{figure}[tb]
    \centering
    \includegraphics[width=0.9\linewidth]{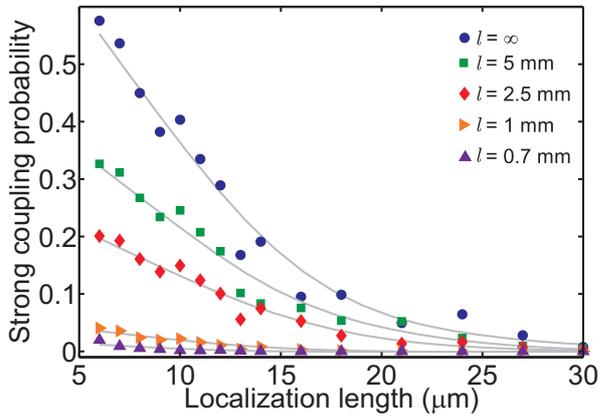}
    \caption{(color online). Total probability of achieving strong coupling plotted as a function of localization length for different loss lengths and for a sample length of $L=\unit{100}{\micro\meter}$. The solid lines are guides to the eye. }
    \label{fig:couplingh}
\end{figure}

Based on the above criterion and the calculated distributions we can estimate the probability that a quantum dot  within a disordered PhC waveguide will be strongly coupled to an Anderson-localized mode. This finite 'yield' of cavities in the strong-coupling regime originates from the statistical nature of multiple scattering, but the outcome is in fact not different from the case of engineered cavities where fabrication imperfections need to be eliminated as good as possible. The resulting strong coupling yield in engineered cavities has to our knowledge not been studied systematically in the literature, but it is not unlikely that the performance of Anderson-localized cavities is superior to standard cavities. We emphasize that in the estimates below it is not assumed that the emitter is optimally positioned at an antinode of an Anderson-localized modes, which corresponds to the realistic experimental situation where no spatial control of quantum dot position is implemented. We assume a transition dipole moment of $d=\unit{0.64}{e\usk\nano\meter}$, which was recently measured for InAs quantum dots  \cite{Johansen2008}. The results are summarized in Fig.~\ref{fig:couplingh} displaying the strong coupling probability as a function of localization length and different loss lengths for a quantum dot located in the center of a sample and tuned into resonance with the nearest Anderson-localized mode present. For a localization length of $\xi>\unit{30}{\micro\meter}$ and negligible loss the fraction of strongly coupled systems is below 1\%. This probability dramatically increases with decreasing localization length leading to 55\% for $\xi=\unit{7}{\micro\meter}$. The rapid increase is mainly due to the drastic increase in the $Q$-factors while the mode volume variations are relatively modest. The presence of loss results in modifications of the \Q-factor distributions (see Fig.~\ref{fig:Q_dist}), and thereby a reduction of the strong-coupling probability, which is investigated systematically in Fig.~\ref{fig:couplingh}.  For $\xi = \unit{7}{\micro\meter}$ and $l=\unit{0.7}{\milli\meter}$ corresponding to the values observed experimentally in the very first experiments on  disordered photonic crystal waveguides with embedded emitters\cite{Stephan}, a strong-coupling probability  of 1\% is predicted. Consequently the observation of strong coupling of a single quantum dot to an Anderson-localized modes appears within experimental reach with disordered PhC waveguides presently available, and finding methods to reduce the localization length and increase the loss length even further, e.g., by inducing correlated disorder, will further increase this probability. In order to estimate the fundamental boundaries set by out-of-plane scattering in a PhC waveguide we consider the simulations of Ref. \cite{Mazoyer2010} for a fixed localization length of $\xi = \unit{7}{\micro\meter}=27a$ and realistic random perturbations in the hole diameter of $\sigma=0.005a$. In this case the loss length can be longer than $l \sim \unit{2.5}{\milli\meter}$ leading to a predicted strong coupling probability above 20$\%$ (Fig.~\ref{fig:couplingh}(b)). We note that in a more complete model, a distribution of out-of-plane light leakage scattering rates is included \cite{Stephan,Savona} as opposed to the single-parameter model employed here, but our conclusions are expected to be robust to such extensions \cite{Baron}. We conclude that apart from providing a new route to cavity QED, Anderson localization may be competitive with engineered nanocavities that inherently suffer from the deteriorating role that disorder plays in this case.

In summary, we have developed a theoretical model for a quantum dot coupled to an Anderson-localized mode. The probability of achieving strong coupling between a single photon and a quantum dot is evaluated and the dependence on the localization length and loss length is studied. Our results show that light-matter entanglement is within experimental reach in a disordered photonic medium, and that Anderson localized random cavities may be competitive with engineered nanocavities for cavity QED experiments.

We gratefully acknowledge  financial support from the Villum Foundation,
the Danish Council for Independent Research (Natural Sciences and Technology and Production
Sciences) and the European Research Council (ERC consolidator grant). We thank Philip Kristensen, Niels Gregersen and Asger Mortensen for useful discussions.


\begin{thebibliography}{99}
\bibitem{Microcavities} A. Kavokin, \textit{et al.}, \textit{Microcavities}, Oxford University Press (2007).
\bibitem{Kimble} H.J. Kimble, Nature \textbf{453}, 1023 (2008).
\bibitem{Vahala} K.J. Vahala, Nature \textbf{424}, 839 (2003).
\bibitem{JMG} J.-M. G\'erard, Topics Appl. Phys. \textbf{90}, 269 (2003).
\bibitem{Yoshie} T. Yoshie, \textit{et al.}, Nature \textbf{432}, 200 (2004).
\bibitem{Pin Sheng} P. Sheng, \textit{Introduction to Wave Scattering, Localization, and Mesoscopic Phenomena.}, Academic Press, San Diego, (1995).
\bibitem{VollmerPRL} J. Topolancik, B. Ilic, F. Vollmer. Phys. Rev. Lett. \textbf{99}, 253901 (2007).
\bibitem{Vollmer} J. Topolancik, \textit{et al.}, Opt. Express \textbf{17}, 12470 (2009).
\bibitem{David} P.D. Garc\'{\i}a, S. Smolka, S. Stobbe, P. Lodahl, Phys. Rev. B. \textbf{82}, 165103 (2010).
\bibitem{Science paper} L. Sapienza, \textit{et al.}, Science \textbf{327}, 1352 (2010).
\bibitem{Deych2001} L.I. Deych, A. Yamilov, A.A. Lisyansky, Phys. Rev. B \textbf{64}, 024201 (2001).
 \bibitem{Skipetrov} S.E. Skipetrov and R. Maynard, Phys. Rev. B \textbf{62}, 886 (2000).
  \bibitem{Tiggelen} B.A. van Tiggelen and S.E. Skipetrov, Phys. Rev. E \textbf{73}, 045601 (2006).
\bibitem{Stephan} S. Smolka, \textit{et al.}, New J. Phys. \textbf{13}, 063044 (2011)
\bibitem{Baron} A. Baron, S. Mazoyer, W. Smigaj, and P. Lalanne, Phys. Rev. Lett. \textbf{107}, 153901 (2011).
\bibitem{Toke-PRL} T. Lund-Hansen, \textit{et al.}, Phys. Rev. Lett. \textbf{101}, 113903 (2008).
\bibitem{Thyrrestrup} H. Thyrrestrup, L. Sapienza, and P. Lodahl, Appl. Phys. Lett. \textbf{96}, 231106 (2010).
\bibitem{NanoOptics} L. Novotny, B. Hecht, \textit{Principles of Nano-Optics}, Cambridge University Press (2006).
\bibitem{selfconsistent} H. Schomerus, M. Titov, P.W. Brouwer,C.W.J. Beenakker, Phys. Rev. B \textbf{65}, 121101 (2002).
\bibitem{MPB} MIT Photonic-Bands (MPB) package is a free program for computing the band structures (dispersion relations) and electromagnetic modes of periodic dielectric structures.
\bibitem{Noda} Y. Akahane, \textit{et al.}, Nature \textbf{425}, 944 (2003).
\bibitem{Lagendijk50Years} A. Lagendijk, B. van Tiggelen, D.S. Wiersma, Phys. Today \textbf{62}, 24. (2009).
\bibitem{Pinheiro2008} F.A. Pinheiro, Phys. Rev. A \textbf{78}, 023812 (2008).
\bibitem{Patra2003} M. Patra, Phys. Rev. E \textbf{67}, 016603 (2003).
\bibitem{Johansen2008} J. Johansen, \textit{et al.}, Phys. Rev. B \textbf{77}, 073303 (2008).
\bibitem{Mazoyer2010} S. Mazoyer, J.P. Hugonin, P. Lalanne, Phys. Rev. Lett. \textbf{103}, 063903 (2009).
\bibitem{Savona} V. Savona, Phys. Rev. B \textbf{83}, 085301 (2011).
\end{thebibliography}
\end{document}